\documentclass[a4paper,11pt]{article}
\usepackage{pos}
\usepackage{float}
\usepackage[backend=bibtex, natbib=true,style=numeric,sorting=none,backref=true]{biblatex}
\usepackage{lineno}

\title{A theoretical model of an off-axis GRB jet}
 \ShortTitle{A theoretical model of an off-axis GRB jet}

\author*[a,b]{B. Betancourt Kamenetskaia}
\author[c]{N. Fraija}
\author[d,e,f,g]{M. Dainotti}
\author[c]{A. Gálvan-Gámez}
\author[h]{R. Barniol Duran}
\author[i,j]{S. Dichiara}

\affiliation[a]{TUM Physics Department, Technical University of Munich, James-Franck-Straße, 85748 Garching, Germany}
\affiliation[b]{LMU Physics Department, Ludwig Maxmillians University, Theresienstr. 37, 80333 Munich, Germany}
\affiliation[c]{Instituto de Astronom\'ia, Universidad Nacional Aut\'onoma de M\'exico, Ciudad de México, México}
\affiliation[d]{Physics Department, Stanford University, 382 Via Pueblo Mall, Stanford, USA}
\affiliation[e]{Space Science Institute, Boulder, CO, USA}
\affiliation[f]{Obserwatorium Astronomiczne, Uniwersytet Jagielloński, ul. Orla 171, 31-501 Kraków, Poland}
\affiliation[g]{Interdisciplinary Theoretical \& Mathematical Science Program, RIKEN(iTHEMS), 2-1 Hirosawa, Wako, Saitama, 351-0198, Japan}
\affiliation[h]{Department of Physics and Astronomy, California State University, Sacramento, 6000 J Street, Sacramento, CA 95819-6041, USA}
\affiliation[i]{Department of Astronomy, University of Maryland, College Park, MD 20742-4111, USA}
\affiliation[j]{Astrophysics Science Division, NASA Goddard Space Flight Center, 8800 Greenbelt Road, Greenbelt, MD 20771, USA}

\emailAdd{boris\_betancourt@ciencias.unam.mx}
\emailAdd{nifraija@astro.unam.mx}

\abstract{In light of the most recent observations of late afterglows produced by the merger of compact objects or by the core-collapse of massive dying stars, we research the evolution of the afterglow produced by an off-axis top-hat jet and its interaction with a surrounding medium. The medium is parametrized by a power law distribution of the form $n(r)\propto r^{-k}$is the stratification parameter and contains the development when the surrounding density is constant ($k=0$) or wind-like ($k=2$). We develop an analytical synchrotron forward-shock model when the outflow is viewed off-axis, and it is decelerated by a stratified medium. Using the X-ray data points collected by a large campaign of orbiting satellites and ground telescopes, we have managed to apply our model and fit the X-ray spectrum of the GRB afterglow associated to SN 2020bvc with conventional parameters. Our model predicts that its circumburst medium is parametrized by a power law with stratification parameter $k=1.5$}

\FullConference{37$^{\rm{th}}$ International Cosmic Ray Conference (ICRC 2021)\\
		July 12th -- 23rd, 2021\\
		Online -- Berlin, Germany}

\begin{document}

\def\aj{AJ}
\def\actaa{Acta Astron.}
\def\araa{ARA\&A}
\def\apj{The Astrophysical Journal}
\def\apjl{ApJ}
\def\apjs{ApJS}
\def\ao{Appl.~Opt.}
\def\apss{Ap\&SS}
\def\aap{A\&A}
\def\aapr{A\&A~Rev.}
\def\aaps{A\&AS}
\def\azh{AZh}
\def\baas{BAAS}
\def\bac{Bull. astr. Inst. Czechosl.}
\def\caa{Chinese Astron. Astrophys.}
\def\cjaa{Chinese J. Astron. Astrophys.}
\def\icarus{Icarus}
\def\jcap{Journal of Cosmology and Astroparticle Physics}
\def\jrasc{JRASC}
\def\mnras{MNRAS}
\def\memras{MmRAS}
\def\na{New A}
\def\nar{New A Rev.}
\def\pasa{PASA}
\def\pra{Phys.~Rev.~A}
\def\prb{Phys.~Rev.~B}
\def\prc{Phys.~Rev.~C}
\def\prd{Phys.~Rev.~D}
\def\pre{Phys.~Rev.~E}
\def\prl{Physical Review Letters}
\def\pasp{PASP}
\def\pasj{PASJ}
\def\qjras{QJRAS}
\def\rmxaa{Rev. Mexicana Astron. Astrofis.}
\def\skytel{S\&T}
\def\solphys{Sol.~Phys.}
\def\sovast{Soviet~Ast.}
\def\ssr{Space~Sci.~Rev.}
\def\zap{ZAp}
\def\nat{Nature}
\def\iaucirc{IAU~Circ.}
\def\aplett{Astrophys.~Lett.}
\def\apspr{Astrophys.~Space~Phys.~Res.}
\def\bain{Bull.~Astron.~Inst.~Netherlands}
\def\fcp{Fund.~Cosmic~Phys.}
\def\gca{Geochim.~Cosmochim.~Acta}
\def\grl{Geophys.~Res.~Lett.}
\def\jcp{J.~Chem.~Phys.}
\def\jgr{J.~Geophys.~Res.}
\def\jqsrt{J.~Quant.~Spec.~Radiat.~Transf.}
\def\memsai{Mem.~Soc.~Astron.~Italiana}
\def\nphysa{Nucl.~Phys.~A}
\def\physrep{Phys.~Rep.}
\def\physscr{Phys.~Scr}
\def\planss{Planet.~Space~Sci.}
\def\procspie{Proc.~SPIE}
\let\astap=\aap
\let\apjlett=\apjl
\let\apjsupp=\apjs
\let\applopt=\ao

\maketitle
\section{Introduction}

\noindent On February 4, 2020, SN 2020bvc was first detected by the All Sky Automated Survey for SuperNovae (ASAS-SN) survey. It was associated to the host galaxy UGC 09379, with a redshift of $z=0.025235$ ($D=120\ \mathrm{Mpc}$)\citep{TNSDR2020381}. A later report confirmed this association and redshift and based on the blue featureless continuum and the absolute magnitude at the discovery of -18.1 classified this event as a young core-collapse supernova \citep{TNSDR2020403}.

Twelve days after its discovery, on February 16, the Very Large Array (VLA) observed the position of SN 2020bvc and detected a point source with a flux density of $66\ \mu\textrm{Jy}$ in the X-band and a luminosity of $1.3\times10^{27}\ {\rm erg\, s^{-1} Hz^{-1}}$ \citep{TNSAN202042}. Such X-ray emission can be seen as another argument in favor of the off-axis GRB scenario, as the radiation can be seen as the afterglow component of a GRB.

Chevalier \citep{1982ApJ...258..790C} explored how an adiabatic flow interacts in a circumstellar density profile of the form $n(r)\propto r^{-2}$ for Type II supernovae. Since then, such a power law has become standard and it has been applied in subsequent publications for other sorts of supernovae, including those by \cite{2004MNRAS.354L..13K}, \cite{1984ApJ...285L..63C}, \cite{2006ApJ...651.1005S} and \cite{1996ApJ...472..257B}, amongst others.  Nevertheless, in a more recent study by Moriya and Tominaga \citep{2012ApJ...747..118M}, the authors demonstrated that the density slope of the surrounding dense wind can explain the variety in the spectrum of Type II luminous supernovae. To this end, they put forward a wind density power law distribution in the form $n(r)\propto r^{-k}$, where $k$ is known as the stratification parameter. They discovered that the ratio of the diffusion timescale in the optically thick region of the wind and the shock propagation timescale after the shock breakout is strongly influenced by the stratification parameter, resulting in variations in the supernova's spectral development. 

In this proceedings, we base ourselves on external forward shocks to introduce an analytic model for calculating the synchrotron emission from an off-axis jet, which is decelerated in an arbitrary orientation with respect to the observer. We analyze the behaviour of our model and compare it with the latest observations of the afterglow of SN2020bvc.

\section{Theoretical Model}\label{Section1}
\subsection{Relativistic Phase: Before the Jet Break}
\noindent Once the outflow launched by the merger of NSs sweeps up enough  cirumburst material, electrons are cooled down by synchrotron radiation. We consider that the jet concentrated within an opening angle $\theta_j$  ``top-hat jet" producing the afterglow emission is not aligned with the observer's line of sight and it differs by an angle $\Delta\theta$.

Considering the adiabatic evolution of the forward shock \citep{1976PhFl...19.1130B, 1997ApJ...489L..37S}, the bulk Lorentz  factor evolves  as 

\begin{equation}\label{Gamma}
    \Gamma\propto(1+z)^{-\frac{k-3}{2}}\xi^{k-3}\zeta_{e}^{2} A_{k}^{-\frac{1}{2}} \theta_{j}^{-1}\Delta\theta^{-(k-3)}\tilde{E}^{\frac{1}{2}}t^{\frac{k-3}{2}} \,,
\end{equation}

where $\tilde{E}$ is the fiducial energy, $\zeta_{e}$ denotes the fraction of electrons that were accelerated by the shock front \citep{2006MNRAS.369..197F} and $\xi\sim1$ is a parameter introduced when the Lorentz factor is approximated as a power law in the radius \citep{1998ApJ...493L..31P}, as is the case in this model. We also denote the redshift as $z$ and the density parameter as $A_{k}$ from the stratification law $n(r)=A_{k}r^{-k}$.

We assume an electron distribution described as $dN/d\gamma_{e}\propto\gamma_{e}^{-p}$ for $\gamma_{e}\geq\gamma_{m}$, where $p$ is the index of the electron distribution and $\gamma_{m}$ is the Lorentz factor of the lowest-energy electrons. Using the bulk Lorentz factor (eq. \ref{Gamma}), the evolution of the minimum and cooling electron Lorentz factors $\gamma_m \propto t^{\frac{k-3}{2}}$ and $\gamma_c$, respectively, and the comoving magnetic field $B'\propto t^{-\frac{3}{2}}$,  the synchrotron spectral breaks and the maximum flux can be written as

\begin{equation}\label{En_br_syn_ism}
    \begin{aligned}
    \nu^{\rm syn}_{\rm m}&\propto (1+z)^{\frac{4-k}{2}}\xi^{k-6}\zeta_{e}^{-2} A_{k}^{-\frac{1}{2}} \epsilon_{e}^2 \epsilon_{B}^{\frac{1}{2}}\theta_{j}^{-2}\Delta\theta^{4-k}\tilde{E}t^{\frac{k-6}{2}}\\
    \nu^{\rm syn}_{\rm c}&\propto  (1+z)^{-\frac{k+4}{2}}\xi^{k+2} A_{k}^{-\frac{1}{2}} (1+Y)^{-2}\epsilon_{B}^{-\frac{3}{2}}\theta_{j}^{2}\Delta\theta^{-(k+4)}\tilde{E}^{-1} t^{\frac{k+2}{2}}\\ 
    F^{\rm syn}_{\rm max} &\propto (1+z)^{\frac{5k-8}{2}}\xi^{12-5k}\zeta_{e} A_{k}^{\frac{5}{2}} \epsilon_{B}^{\frac{1}{2}}D_{z}^{-2}\theta_{j}^{2}\Delta\theta^{5k-18}\tilde{E}^{-1} t^{\frac{12-5k}{2}}\,,
    \end{aligned}
\end{equation}

where where $Y$ is the Compton parameter, $D_{z}$ is the luminosity distance, $\epsilon_{e}$ is the fraction of the shock's thermal energy density that is transmitted to the electrons and $\epsilon_{B}$ is the fraction turned into magnetic energy density \citep{2015ApJ...804..105F}.

Using the synchrotron spectral breaks, the maximum flux (eqs. \ref{En_br_syn_ism}), the synchrotron light curves in the fast- and slow-cooling regimes can be written  as:

\begin{equation}\label{Fl_syn_ism_fc}
    F^{\rm syn}_{\nu}\propto
    \begin{cases}
    t^{\frac{17-8k}{3}}  \, \nu^{\frac13},\hspace{1.7cm} \nu < \nu^{\rm syn}_{\rm c},\cr
    t^{\frac{26-9k}{4}}\,\nu^{-\frac{1}{2}}\,\hspace{1.6cm}  \nu^{\rm syn}_{\rm c}<\nu<\nu^{\rm syn}_{\rm m},\,\,\,\,\,\cr
    t^{\frac{32-10k-6p+kp}{4}}\,\nu^{-\frac{p}{2}},\hspace{0.5cm}\nu^{\rm syn}_{\rm m}<\nu\,. \cr
    \end{cases}
\end{equation}

and

\begin{equation}\label{Fl_syn_ism_sc}
    F^{\rm syn}_{\nu}\propto
    \begin{cases}
    t^{\frac{21-8k}{3}}  \, \nu^{\frac13},\hspace{2.4cm} \nu<\nu^{\rm syn}_{\rm m},\cr
    t^{\frac{(p-11)k-6(p-5)}{4}}\nu^{-\frac{p-1}{2}},\hspace{0.8cm} \nu^{\rm syn}_{\rm m}<\nu<\nu^{\rm syn}_{\rm c},\,\,\,\,\,\cr
    t^{\frac{32-10k-6p+kp}{4}} \,\nu^{-\frac{p}{2}},\hspace{1.1cm}\nu^{\rm syn}_{\rm c}<\nu\,. \cr
    \end{cases}
\end{equation}

It is worth noting that when $k=0$, the observable quantities derived in \cite{1998ApJ...497L..17S} and the light curves of the synchrotron forward-shock emission are recovered \citep[e.g., see][]{2016ApJ...831...22F,2016ApJ...818..190F}.\\ 

\subsection{Lateral expansion: After the jet break}

During the lateral expansion phase, the beaming cone of the radiation emitted off-axis,  $\Delta \theta$, broaden increasingly until this cone reaches our field of view $\Gamma_{\rm j}\sim \Delta \theta^{-1}$ \citep{2002ApJ...570L..61G,  2017arXiv171006421G}.   Considering the adiabatic evolution in this phase,  the bulk Lorentz  factor evolves as:

\begin{equation}\label{Gamma_l_ism}
    \Gamma\propto(1+z)^{\frac{1}{2}}\xi^{-1} A_{k}^{\frac{1}{2(k-3)}} \tilde{E}^{-\frac{1}{2(k-3)}}t^{-\frac{1}{2}}   \,.
\end{equation}

Using the bulk Lorentz factor (eq. \ref{Gamma_l_ism}) and the synchrotron afterglow theory introduced in \cite{1999ApJ...519L..17S}, the synchrotron spectral breaks and the maximum flux evolve as:

\begin{equation}\label{En_br_syn_ism_l}
    \begin{aligned}
    \nu^{\rm syn}_{\rm m}&\propto (1+z)\xi^{-4}\zeta_{e}^{-2}A_{k}^{\frac{1}{2(k-3)}}\epsilon_{e}^2 \epsilon_{B}^{\frac{1}{2}}\tilde{E}^{\frac{k-4}{2(k-3)}}t^{-2}\cr
    \nu^{\rm syn}_{\rm c}&\propto  (1+z)^{-1}\xi^{0}A_{k}^{\frac{5}{2(k-3)}}\epsilon_{B}^{-\frac{3}{2}}(1+Y)^{-2}\tilde{E}^{\frac{4-3k}{2(k-3)}}t^{0}\cr 
    F^{\rm syn}_{\rm max} &\propto (1+z)^{3}\xi^{-2}\zeta_{e}A_{k}^{\frac{1}{2(3-k)}}\epsilon_{B}^{\frac{1}{2}}D_{z}^{-2}\tilde{E}^{\frac{8-3k}{2(3-k)}}t^{-1}\,.
    \end{aligned}
\end{equation}

Using the observed synchrotron spectrum in the slow-cooling regime with eq. (\ref{En_br_syn_ism_l}), the synchrotron light curves in the slow-cooling regimes can be written  as:

\begin{equation}\label{AfterSlowCooling2}
    F^{\rm syn}_{\nu}\propto
    \begin{cases}
    t  \, \nu^{\frac13},\hspace{1.8cm} \nu<\nu^{\rm syn}_{\rm m},\cr
    t^{-p}\nu^{-\frac{p-1}{2}},\hspace{1.0cm} \nu^{\rm syn}_{\rm m}<\nu<\nu^{\rm syn}_{\rm c},\,\,\,\,\,\cr
    t^{-p} \,\nu^{-\frac{p}{2}},\hspace{1.2cm}\nu^{\rm syn}_{\rm c}<\nu\,, \cr
    \end{cases}
\end{equation}

\section{Results and discussion}\label{Results}
\begin{figure}[H]
    \centering
    \includegraphics[scale=0.6]{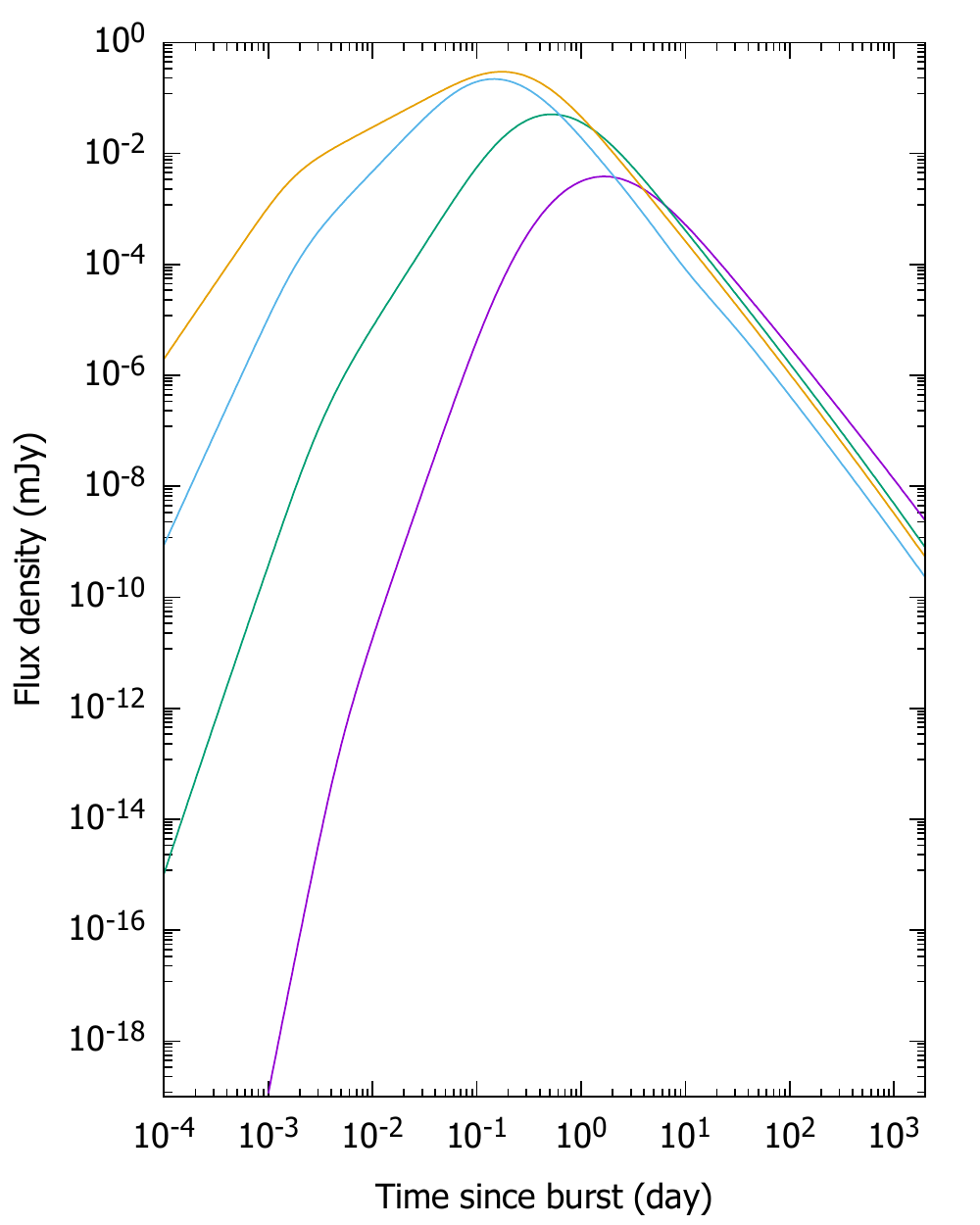}
    \caption{The X-ray light curves for different stratification parameters.}   
    \label{LightCurves}
\end{figure}

In Figure \ref{LightCurves} the light curves in the X-ray band (1 keV) are plotted. Different stratification parameters were used; the purple curve corresponds to $k=0$, the green one to $k=1$, the blue one to $k=1.5$ and the yellow one to $k=2$. The rising behavior of the flux is represented by the relativistic phase, while the decrease is due to the lateral expansion. The Figure shows that the behavior after the jet breaks is independent of the stratification of the medium, while an increase of $k$ in the relativistic phase leads to flatter profiles. The parameters used for this figure are: $\tilde{E}=10^{51}\,{\rm erg}$, $\epsilon_{\rm B}=10^{-2}$, $\epsilon_{\rm e}=10^{-1}$, $p=2.6$, $\zeta_{e}=1$, $\xi=1$, $\Delta\theta=15^{\circ}$, $\theta_{j}=5^{\circ}$ and $D_z=26.5\,{\rm Mpc}$. The density parameters were: $A_{0}=1\ \mathrm{cm}^{-3}$, $A_{1}=1.5\times10^{19}\ \mathrm{cm}^{-2}$, $A_{1.5}=2.7\times10^{28}\ \mathrm{cm}^{-\frac32}$ and $A_{2}=3\times10^{36}\ \mathrm{cm}^{-1}$.

\begin{figure}[H]
    \centering
    \includegraphics{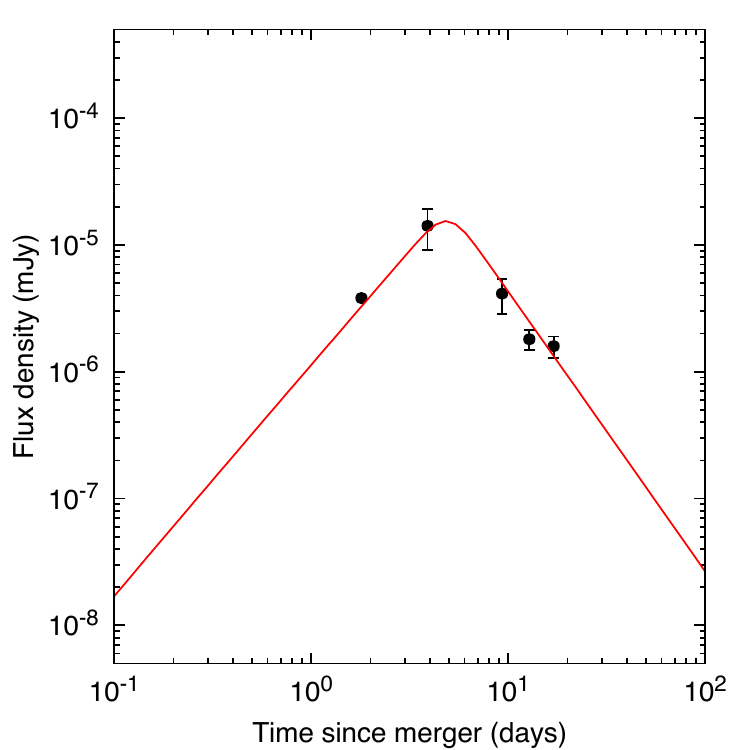}
    \caption{The X-ray data points of SN 2020bvc with the best-fit curve obtained with the model presented in this proceedings for a stratification parameter of $k=1.5$.}   
    \label{Fit_SN2020bvc}
\end{figure}

Figure \ref{Fit_SN2020bvc} shows the X-ray observations of SN 2020bvc with the best fit synchrotron light curve generated by the deceleration of an off-axis jet in a medium with stratification parameter $k=1.5$. The synchrotron light curve is plotted for the following parameter values: $\tilde{E}=5.3\times10^{49}\,{\rm erg}$, $\epsilon_{\rm B}=2\times10^{-2}$ $\epsilon_{\rm e}=3.5\times10^{-3}$, $A_{k}=8.47\times10^{25}\ \mathrm{cm}^{-\frac32}$, $p=2.2$, $\Delta\theta=23^{\circ}$, $\theta_{j}=5^{\circ}$ and $D_z=120\,{\rm Mpc}$.

\section{Conclusion}

Using external forward shocks as our starting point, we have derived a theoretical model to describe the afterglow emission of an off-axis relativistic jet. Influenced by earlier work on supernova emission, we have assumed that the jet interacts with a medium parametrized by a power law number density distribution $\propto R^{-k}$. This general approach is advantageous, as it allows us to not only consider a homogeneous medium ($k=0$) and a wind-like medium ($k=2$), but regions with non-standard stratification parameters, in particular $k=1$, $1.5$ or $2.5$.

Our model has considered the jet's evolution during the relativistic phase and after the jet break. In both stages, we have calculated the synchrotron light curves in the fast- and slow-cooling regimes, which agree with previous literature on the subject for the particular cases of a homogeneous and a wind-like medium. 

We have analyzed the behaviour of the light curves for different sets of parameters. In the case of variation of the stratification parameter, we have noticed that during the relativistic phase, an increase of this parameter leads to flatter profiles. As for the case of the flux after the jet break, we have shown that its time evolution is independent of $k$. Therefore, an advantage of this model is the freedom with which one can explain the early-time evolution of the radiation, without spoiling the long-time results.

We have applied our model to the radiation from SN 2020bvc and have been able to successfully fit all reported values with the parameters specified in Section \ref{Results}. A stratification parameter of $k=1.5$ was used.

\acknowledgments{
We acknowledge the support from Consejo Nacional de Ciencia y Tecnolog\'ia (CONACyT), M\'exico, grants IN106521.

}

\printbibliography


\end{document}